\documentclass[]{ws-ijmpa}
\usepackage{amsmath}
\usepackage{amssymb}

\begin{document}

\title{Topological quantum D-branes and wild embeddings from exotic smooth
$\mathbb{R}^{4}$}

\author{Torsten Asselmeyer-Maluga}%
\address{German Aerospace center, Rutherfordstr. 2, 12489 Berlin \\ torsten.asselmeyer-maluga@dlr.de%
} 
\author{Jerzy Kr\'ol}%
\address{University of Silesia, Institute of Physics, ul. Uniwesytecka 4, 40-007
Katowice \\ iriking@wp.pl %
}

\maketitle
\begin{history}
\received{Day Month Year}
\revised{Day Month Year}
\end{history}

\begin{abstract}
This is the next step of uncovering the relation between string theory
and exotic smooth $\mathbb{R}^{4}$. Exotic smoothness of $\mathbb{R}^{4}$
is correlated with D6 brane charges in IIA string theory. We construct
wild embeddings of spheres and relate them to a class of topological
quantum D$p$-branes as well to KK theory. These branes emerge when
there are non-trivial NS-NS $H$-fluxes where the topological classes
are determined by wild embeddings $S^{2}\to S^{3}$. Then wild embeddings
of higher dimensional $p$-complexes into $S^{n}$ correspond to D$p$-branes.
These wild embeddings as constructed by using gropes are basic objects
to understand exotic smoothness as well Casson handles. Next we build
$C^{\star}$-algebras corresponding to the embeddings. Finally we
consider topological quantum D-branes as those which emerge from wild
embeddings in question. We construct an action for these quantum D-branes
and show that the classical limit agrees with the Born-Infeld action
such that flat branes = usual embeddings.

\end{abstract}

\section{Introduction}

Despite the substantial effort toward quantizing gravity in 4 dimensions,
this issue is still open. One of the best candidates till now is the
superstring theory formulated in 10 dimensions. A way from superstring
theory to 4-dimensional quantum gravity or standard model of particle
physics (minimal supersymmetric extension thereof) is, at best, highly
nonunique. Many techniques of compactifications and flux stabilization
along with specific model-building branes configurations and dualities,
were worked out toward this end within the years. Possibly some important
data of a fundamental character are still missing enabling the connection
with physics in dimension 4. 

In this paper we follow the idea from \cite{AsselmeyerKrol2011,AssKrol2010ICM}
that different smoothings of Euclidean $\mathbb{R}^{4}$ are presumably
crucial for the program of QG and string theory. These structures
are footed certainly in dimension 4 and have great importance to physics
\cite{AsselmeyerKrol2010,AsselmeyerKrol2009,Bra:94b,BraRan:93,Sladkowski2001}.
Here we again try to consider exotic $\mathbb{R}^{4}$'s as serving
a link between higher dimensional superstring theory and 4-dimensional
,,physical'' theories. String theory D- and NS-branes in some backgrounds
are correlated naturally with exotic smoothness on $\mathbb{R}^{4}$
appearing in these backgrounds \cite{AsselmeyerKrol2011}. Moreover,
when taking quantum limit of D-branes and spaces, such that these
become represented by separable $C^{\star}$-algebras, the connection
with exotic $\mathbb{R}^{4}$'s extends naturally. This is due to
the representing exotic $\mathbb{R}^{4}$'s by convolution $C^{\star}$-algebras
of the codimension-one foliations of certain 3-sphere. In this paper
we focus on the topological level underlying the quantum branes and
exotic $\mathbb{R}^{4}$'s connection. We show that there exists topological
counterparts of D-branes in a quantum regime. Namely, by the use of
$C^{*}$algebra approach to quantum D-branes the manifold model of
a quantum D-brane as wild embedding is constructed. Then we show that
the $C^{*}$algebra of the wild embedding is isomorphic to the $C^{*}$algebra
of the quantum D-brane. We call the wild embeddings representing quantum
D-branes as \emph{topological quantum branes}. Moreover, the low dimensional
wild embedding, i.e. $S^{2}\to S^{3}$ expresses the existence of
the non-trivial $B$-field on the quantum level. Next we construct
a quantum version of an action using cyclic cohomology of $C^{\star}$-algebra.
In the classical limit this action reduces to the Born-Infeld one
for flat branes given by tame embedding.

The basic technical ingredient of the analysis of small exotic $\mathbb{R}^{4}$'s
enabling uncovering many applications also in string theory is the
relation between exotic (small) $\mathbb{R}^{4}$'s and non-cobordant
codimension-1 foliations of the $S^{3}$ as well gropes and wild embeddings
as shown in \cite{AsselmeyerKrol2009}. The foliation are classified
by the Godbillon-Vey class as element of the cohomology group $H^{3}(S^{3},\mathbb{R})$.
By using the $S^{1}$-gerbes it was possible to interpret the integral
elements $H^{3}(S^{3},\mathbb{Z})$ as characteristic classes of a
$S^{1}$-gerbe over $S^{3}$ \cite{AsselmeyerKrol2009a}.

\section{Small exotic $\mathbb{R}^{4}$, gropes and foliations\label{sec:Small-exotic-R4-gropes}}

In this short section we will only give a rough overview about a relation
between small exotic $\mathbb{R}^{4}$ and foliations. Some of the
details can be found in \cite{AsselmeyerKrol2009} and more detailed
approach will appear here \cite{AsselmeyerKrol2011a}. At first we
will start with some facts about exotic 4-spaces. 

An exotic $\mathbb{R}^{4}$ is a topological space with $\mathbb{R}^{4}-$topology
but with a different (i.e. non-diffeomorphic) smoothness structure
than the standard $\mathbb{R}_{std}^{4}$ getting its differential
structure from the product $\mathbb{R}\times\mathbb{R}\times\mathbb{R}\times\mathbb{R}$.
The exotic $\mathbb{R}^{4}$ is the only Euclidean space $\mathbb{R}^{n}$
with an exotic smoothness structure. The exotic $\mathbb{R}^{4}$
can be constructed in two ways: by the failure to arbitrarily split
a smooth 4-manifold into pieces (large exotic $\mathbb{R}^{4}$) and
by the failure of the so-called smooth h-cobordism theorem (small
exotic $\mathbb{R}^{4}$). Here we will use the second method. 

Consider the following situation: one has two topologically equivalent
(i.e. homeomorphic), simple-connected, smooth 4-manifolds $M,M'$,
which are not diffeomorphic. There are two ways to compare them. First
one calculates differential-topological invariants like Donaldson
polynomials \cite{DonKro:90} or Seiberg-Witten invariants \cite{Akb:96}.
But there is another possibility: It is known that one can change
a manifold $M$ to $M'$ by using a series of operations called surgeries.
This procedure can be visualized by a 5-manifold $W$, the cobordism.
The cobordism $W$ is a 5-manifold having the boundary $\partial W=M\sqcup M'$.
If the embedding of both manifolds $M,M'$ in to $W$ induces homotopy-equivalences
then $W$ is called an h-cobordism. Furthermore we assume that both
manifolds $M,M'$ are compact, closed (no boundary) and simply-connected.
As Freedman \cite{Fre:82} showed a h cobordism implies a homeomorphism,
i.e. h-cobordant and homeomorphic are equivalent relations in that
case. Furthermore, for that case the mathematicians \cite{CuFrHsSt:97}
are able to prove a structure theorem for such h-cobordisms:\\
 \emph{Let $W$ be a h-cobordism between $M,M'$. Then there are
contractable submanifolds $A\subset M,A'\subset M'$ together with
a sub-cobordism $V\subset W$ with $\partial V=A\sqcup A'$, so that
the h-cobordism $W\setminus V$ induces a diffeomorphism between $M\setminus A$
and $M'\setminus A'$.} \\
 Thus, the smoothness of $M$ is completely determined (see also
\cite{Akbulut08,Akbulut09}) by the contractible submanifold $A$
and its embedding $A\hookrightarrow M$ determined by a map $\tau:\partial A\to\partial A$
with $\tau\circ\tau=id_{\partial A}$ and $\tau\not=\pm id_{\partial A}$($\tau$
is an involution). One calls $A$, the \emph{Akbulut cork}. According
to Freedman \cite{Fre:82}, the boundary of every contractible 4-manifold
is a homology 3-sphere. This theorem was used to construct an exotic
$\mathbb{R}^{4}$. Then one considers a tubular neighborhood of the
sub-cobordism $V$ between $A$ and $A'$. The interior $int(V)$
(as open manifold) of $V$ is homeomorphic to $\mathbb{R}^{4}$. If
(and only if) $M$ and $M'$ are homeomorphic, but non-diffeomorphic
4-manifolds then $int(V)\cap M$ is an exotic $\mathbb{R}^{4}$. As
shown by Bizaca and Gompf \cite{Biz:94a,BizGom:96} one can use $int(V)$
to construct an explicit handle decomposition of the exotic $\mathbb{R}^{4}$.
We refer for the details of the construction to the papers or to the
book \cite{GomSti:1999}. The idea is simply to use the cork $A$
and add some Casson handle $CH$ to it. The interior of this construction
is an exotic $\mathbb{R}^{4}$. Therefore we have to consider the
Casson handle and its construction in more detail. Briefly, a Casson
handle $CH$ is the result of attempts to embed a disk $D^{2}$ into
a 4-manifold. In most cases this attempt fails and Casson \cite{Cas:73}
looked for a substitute, which is now called a Casson handle. Freedman
\cite{Fre:82} showed that every Casson handle $CH$ is homeomorphic
to the open 2-handle $D^{2}\times\mathbb{R}^{2}$ but in nearly all
cases it is not diffeomorphic to the standard handle \cite{Gom:84,Gom:89}.
The Casson handle is built by iteration, starting from an immersed
disk in some 4-manifold $M$, i.e. a map $D^{2}\to M$ with injective
differential. Every immersion $D^{2}\to M$ is an embedding except
on a countable set of points, the double points. One can kill one
double point by immersing another disk into that point. These disks
form the first stage of the Casson handle. By iteration one can produce
the other stages. Finally consider not the immersed disk but rather
a tubular neighborhood $D^{2}\times D^{2}$ of the immersed disk,
called a kinky handle, including each stage. The union of all neighborhoods
of all stages is the Casson handle $CH$. So, there are two input
data involved with the construction of a $CH$: the number of double
points in each stage and their orientation $\pm$. Thus we can visualize
the Casson handle $CH$ by a tree: the root is the immersion $D^{2}\to M$
with $k$ double points, the first stage forms the next level of the
tree with $k$ vertices connected with the root by edges etc. The
edges are evaluated using the orientation $\pm$. Every Casson handle
can be represented by such an infinite tree. 

The main idea of the relation between small exotic $\mathbb{R}^{4}$
is the usage of a radial family of small exotic $\mathbb{R}^{4}$,
i.e. a continuous family of exotic $\{\mathbb{R}_{\rho}^{4}\}_{\rho\in[0,+\infty]}$
with parameter $\rho$ so that $\mathbb{R}_{\rho}^{4}$ and $\mathbb{R}_{\rho'}^{4}$
are non-diffeomorphic for $\rho\not=\rho'$. This radial family has
a natural foliation (see Theorem 3.2 in \cite{DeMichFreedman1992}
which can be induced by a polygon $P$ in the two-dimensional hyperbolic
space $\mathbb{H}^{2}$. The area of $P$ is a well-known invariant,
theGodbillon-Vey class as element in $H^{3}(S^{3},\mathbb{R})$, determing
a codimension-one foliation on the 3-sphere (firstly constructed by
Thurston \cite{Thu:72}, see also the book \cite{Tamura1992} chapter
VIII for the details). This 3-sphere is part of the boundary $\partial A$
of the Akbulut cork $A$ (or better there is an emebdding $S^{3}\to\partial A$).
Furthermore one can show that the codimension-one foliation of the
3-sphere induces a codimension-one foliation of $\partial A$ so that
the area of the corresponding polygons (and therefore the invariants)
agree. The Godbillon-Vey invariant $[GV]\in H^{3}(S^{3},\mathbb{R})$
of the foliation is related to the parameter of the radial family
by $\left\langle GV,[S^{3}]\right\rangle =\rho^{2}$ using the pairing
between cohomology and homology (the fundamental class $[S^{3}]\in H_{3}(S^{3})$).

Thus we are able to obtain a relation between an exotic $\mathbb{R}^{4}$
(of Bizaca as constructed from the failure of the smooth h-cobordism
theorem) and codimension-one foliation of the $S^{3}$. Two non-diffeomorphic
exotic $\mathbb{R}^{4}$implying non-cobordant codimension-one foliations
of the 3-sphere described by the Godbillon-Vey class in $H^{3}(S^{3},\mathbb{R})$
(proportional to the are of the polygon). This relation is very strict,
i.e. if we change the Casson handle then we must change the polygon.
But that changes the foliation and vice verse. Finally we obtained
the result:\\
\emph{The exotic $\mathbb{R}^{4}$ (of Bizaca) is determined by
the codimension-1 foliations with non-vanishing Godbillon-Vey class
in $H^{3}(S^{3},\mathbb{R}^{3})$ of a 3-sphere seen as submanifold
$S^{3}\subset\mathbb{R}^{4}$. We say: the exoticness is localized
at a 3-sphere inside the small exotic $\mathbb{R}^{4}$.}

\section{RR charges of D6-Branes in the presence of $B$-field\label{sec:RR-charges-D6-with-B-field}}

In this section, we will describe the direct reference of 4-dimensional
structures to the dynamics of special higher dimensional branes, the
D6-brane, in flat spacetime. This D6-brane is usually involved in
building various ,,realistic'' 4-dimensional models derived from brane
configurations. We will analyze this case separately along with the
discussion of compactifications in string theory in a forthcoming
paper.

Let us consider the D6-brane of IIA string theory in flat 10 dimensional
spacetime and assume a vanishing B-field. The world-volumes of flat
Dp-branes are classified by $K_{1}(\mathbb{R}^{p+1})$ where this
K-homology group is understood as $K^{1}(C_{0}(\mathbb{R}^{p+1}))$.
Then there is the isomorphism $K_{1}(\mathbb{R}^{p+1})=K_{1}(S^{p+1})$
between the K-groups induced by the isomorphism of the reduced $C^{\star}$
algebra of functions $C_{0}(\mathbb{R}^{p+1})=C(S^{p+1})$. Their
charges, constraining the dynamics of the brane, are dually described
by $K^{1}(\mathbb{R}^{9-p})=K^{1}(S^{9-p})$. In the case of the D6-branes,
the group $K^{1}(S^{3})$ classifies the RR charges of flat D6-branes
in flat 10-dimensional spacetime \cite{Witten1998}. 

In case of a non-vanishing \emph{B}-field for a stable D6-brane, the
\emph{B}-field needs be non-trivial on the space $\mathbb{R}^{3}$
transversal to the world-volume (on the brane we have $dB=H$). Hence
the \emph{B}-field must be nontrivial on the space $S^{3}$. It is
convienient to adopt (see \cite{Szabo2008a}, p. 654) the definition
of $B$ - field on a manifold $X$ as a class of gerbe (with connection),
which referes directly to the topologically non-trivial $H^{3}(S^{3},\mathbb{Z})$
classes instead of local 2-form $B$:

\begin{definition}

A $B$-field $(X,H)$ is a gerbe with one-connection over $X$ and
characteristic class $[H]\in H^{3}(X,Z)$ which is an NS\textendash{}NS
H-flux.

\end{definition}

Then the classification of D6-brane charges in IIA type superstring
theory in flat space is influenced by the presence of non-trivial
B-field and now is given by the twisted K-theory $K_{H}(S^{3})$,
so that $K^{1}(S^{3},H)=\mathbb{Z}_{k}$ with $0\neq[H]=k\in H^{3}(S^{3},\mathbb{Z})$.
Hence the dynamics of D6-branes in type IIA superstring theory on
flat spacetime is influenced by a non-zero B-field. 

Following our philosophy already implicitly present in our previous
work, the source of the non-trivial \emph{B}-field on $S^{3}$ (hence
$H\neq0$) is the exoticness of the ambient $\mathbb{R}^{4}$. This
result is motivated by our work that some (small) exotic $\mathbb{R}_{H}^{4}$'s
correspond to non-trivial classes $[H]\in H^{3}(S^{3},\mathbb{Z})$
and conversely, where $S^{3}$ is part of the boundary of the Akbulut
cork \cite{AsselmeyerKrol2009,AsselmeyerKrol2009a}. Moreover, exotic
smoothness of $\mathbb{R}_{H}^{4}$ twists the K-theory groups $K^{\star}(S^{3})$
\cite{AsselmeyerKrol2010} where the 3-sphere $S^{3}$ lies at the
boundary of the Akbulut cork. Hence the dynamics induced by D6-branes
in the spacetime $\mathbb{R}_{H}^{4}\times\mathbb{R}^{5,1}$ is equivalent
to the dynamics induced by a D6-brane with a non-zero B-field on the
transversal $\mathbb{R}^{3}$ compactified to $S^{3}$. Finally we
get: 

\begin{theorem}

RR charges of D6-branes in string theory IIA in the presence of a
non-trivial B-field ($H\neq0$), (these charges are classified by
$K_{H}(S^{3})$, and $[H]\in H^{3}(S^{3},\mathbb{Z})$), are related
to exotic smoothness of small $\mathbb{R}_{H}^{4}$. This exotic $\mathbb{R}_{H}^{4}$
corresponds to $[H]$ which twists $K(S^{3})$ \cite{AsselmeyerKrol2010},
where $S^{3}\subset\mathbb{R}^{4}$ lies at the boundary of the Akbulut
cork and $S^{3}$ is transverse to the branes. Thus, changing the
smoothness of $\mathbb{R}^{4}$ gives rise to the change of the allowed
charges for D6 branes inducing a change of the dynamics. 

\end{theorem}

We saw that the geometric realization of (classical) D-branes in certain
backgrounds of string theory is correlated with small exotic $\mathbb{R}^{4}$'s
which can be all embedded in the standard smooth $\mathbb{R}^{4}$.
As was shown in our previous paper \cite{AsselmeyerKrol2011,AsselmeyerKrol2011b},
quantum D-branes correspond to the net of exotic smooth $\mathbb{R}^{4}$'s
embedded in certain exotic smooth $\mathbb{R}^{4}$. An intriguing
interpretation for this correspondence can be given by: \emph{in some
limit of IIA superstring theory, small exotic smooth $\mathbb{R}^{4}$'s
can be considered as carrying the RR charges of D6 branes}. A generalization
of this concept will be studied in the next section.

\section{D6-brane charges and embeddings of $(4k-1)$- into $6k$-manifolds}

In the usual definition of Dp-branes, one considers embedded $p$-dimensional
objects in some higher-dimensional space. In the presence of NS-NS
H-fluxes, curved (twisted) classical branes are defined still as submanifolds
with some extra topological condition (cancelling the anomaly) (see
\cite{Szabo2008a}, Def. 1.14, p. 654): 

\begin{definition}

Twisted D-brane in a $B$-field $(X,H)$ is a triple $(W,E,\phi)$,
where $\phi:W\to X$ is a closed, embedded oriented submanifold with
$\phi^{\star}H=W_{3}(W)$, and $E\in K0(W)$ 

\end{definition}

where $W_{3}(W)\in H^{3}(W,\mathbb{Z})$ is the third Stiefel-Whithney
class of the normal bundle of $W$ in $X$, $N(X/W)$. This case of
non-trivial H-flux directly referes to the non-commutative geometry
tools hence quantum D-branes are considered naturally here \cite{Szabo2008a}.
It is known \cite{Kapustin1999} that in the presence of a topologically
non-trivial $B$-field world-volumes of D-branes are rather described
as noncommutative spaces in Conne's sense. We try to find a topological
characterisation for D-branes as a kind of embedding also in the quantum
regime. These branes emerge in case of the non-trivial H-flux too,
and we call them \emph{topological quantum branes}. However the reason
for the existence of this H-flux is deeply rooted in the geometry
of some exotic $\mathbb{R}^{4}$. Exotic $\mathbb{R}^{4}$ itself
referes to non-commutative spaces and tools from non-commutative geometry,
which can be seen as one reasons behind the quantum description of topological
branes .

As explained in the previous section, the charges of D-branes are
given by some (twisted) K-theory classes also related to (twisted)
cohomology. In case of a D6-brane, the charge is classified by the
twisted K-theory $K_{H}(S^{3})$ with $[H]\in H^{3}(S^{3},\mathbb{Z})$
(\v{C}ech 3-cocycle). Therefore one has a 3-sphere determinig the
charge of a 6-dimensional object. To simplify the discussion, we will
compactify the (possible) infinite D6-brane to a 6-sphere $S^{6}$
in the following.

Now we start with a short discussion of embeddings $S^{3}\to S^{6}$
as an example $k=1$ of a general map $S^{4k-1}\to S^{6k}$ to understand
the charge classification. As Haefliger \cite{Haefliger1962} showed,
the isotopy classes of these embeddings are determined by the integer
classes (Hopf invariant) in $H^{3}(S^{3},\mathbb{Z})$. Thus the $4k-1$
space is knotted in the $6k$ space. This phenomenon depends strongly
on smoothness, i.e. it disappears for continuous or PL embeddings.
Usually every $n-$sphere or every homology $n-$sphere unknots (in
PL or TOP) in $\mathbb{R}^{m}$ for $m\geq n+3$, i.e. for codimension
$m-n=3$ or higher. Of course, one has the usual knotting phenomena
in codimension $2$ and the codimension $1$ was shown to be unique
for embeddings $S^{n}\to S^{n+1}$ (for $n\geq6$) but is hard to
solve in other cases. 

Let $\Sigma\to S^{6}$ be an embedding of a homology 3-sphere $\Sigma$
(containing the case $S^{3}$). Then the normal bundle of $F$ is
trivial (definition of an embedding) and homotopy classes of trivialisations
of the normal bundle (normal framings) are classified by the homotopy
classes $[\Sigma,SO(3)]$ with respect to some fixed framing. There
is an isomorphism $[\Sigma,SO(3)]=[\Sigma,S^{2}]$ (so-called Pontrjagin-Thom
construction) and $[\Sigma,S^{2}]$ can be identified with $H^{3}(\Sigma,\mathbb{Z})=\mathbb{Z}$.
That is one possible way to get the classification of isotopy classes
of embeddings $\Sigma\to S^{6}$ by elements of $H^{3}(\Sigma,\mathbb{Z})=\mathbb{Z}$.
A class $[H]$ in $H^{3}(\Sigma,\mathbb{Z})$ determines via an injective
homomorphism a (deRham-)cocycle $H\in H^{3}(\Sigma,\mathbb{R})$.
If $H$ is the field strength for the B-field then the 3-form $H$
must be a multiply of the volume form on $\Sigma$ and we have\[
\intop_{\Sigma}H=Q\not=0\]
by the usual pairing between homology and cohomology. By the cellular
approximation theorem, the elements in $H^{3}(\Sigma,\mathbb{Z})$
are determined by $H^{3}(S^{3},\mathbb{Z})$. Combined with our result
that $H^{3}(S^{3},\mathbb{Z})$ determines some exotic $\mathbb{R}^{4}$
we have shown:

\begin{theorem}(The topological origins of the allowed D6-brane charges)

\label{theo:D6-brane-charge}

Let $\mathbb{R}_{H}^{4}$ be some exotic $\mathbb{R}^{4}$ determined
by some 3-form $H$, i.e. by a codimension-1 foliation on the boundary
$\partial A$ of the Akbulut cork $A$. The codimension-1 foliation
on $\partial A$ is determined by $H^{3}(\partial A,\mathbb{R})$.
Each integer class in $H^{3}(\partial A,\mathbb{Z})$ determines the
isotopy class of an embedding $\partial A\to S^{6}$. Hence, the group
of allowed charges of D6-branes in the presence of B-field $H$ ,
i.e. $K_{H}^{\star}(S^{3})$ is determined equivalently by the isotopy
classes of embeddings $\partial A\to S^{6}$. The classes of $H$-field
are topologically determined by the isotopy classes of the embeddings,
which affects the allowed charges of D6-branes. 

\end{theorem}

But more is true. Given two embeddings $F_{i}:\Sigma_{i}\to S^{6}$
between two homology 3-spheres $\Sigma_{i}$ for $i=0,1$. A homology
cobordism is a cobordism between $\Sigma_{0}$ and $\Sigma_{1}$.
This cobordism can be embedded in $S^{6}\times[0,1]$ determining
the homology bordism class of the embedding. Then two embeddings of
an oriented homology 3-sphere in $S^{6}$ are isotopic if and only
if they are homology bordant.

\section{From wild embeddings to topological quantum D-branes}

In this section we try to give a geometric approach to quantm D-branes
using wild embeddings of trivial complexes into $S^{n}$ or $\mathbb{R}^{n}$.
This point of view is supported by the Theorem \ref{theo:D6-brane-charge}
above. Here we will describe a dimension-independent way: every wild
embedding $j$ of a $p-$dimensional complex $K$ into the $n-$dimensional
sphere $S^{n}$ is determined by the fundamental group $\pi_{1}(S^{n}\setminus j(K))$
of the complement. This group is perfect and uniquely representable
by a 2-dimensional complex, a singular disk or grope (see \cite{Can:79}).
As we showed in \cite{AsselmeyerKrol2009}, the exotic $\mathbb{R}^{4}$
is related to a grope. Thus, these constructed \emph{topological quantum}
D-branes are determined by exotic $\mathbb{R}^{4}$'s which act as
a kind of germ for the branes.

\subsection{Wild and tame embeddings\label{sub:Wild-and-tame-embed}}

We call a map $f:N\to M$ between two topological manifolds an embedding
if $N$ and $f(N)\subset M$ are homeomorphic to each other. From
the differential-topological point of view, an embedding is a map
$f:N\to M$ with injective differential on each point (an immersion)
and $N$ is diffeomorphic to $f(N)\subset M$. An embedding $i:N\hookrightarrow M$
is \emph{tame} if $i(N)$ is represented by a finite polyhedron homeomorphic
to $N$. Otherwise we call the embedding \emph{wild}. There are famous
wild embeddings like Alexanders horned sphere or Antoine's necklace.
In physics one uses mostly tame embeddings but as Cannon mentioned
in his overview \cite{Can:78}, one needs wild embeddings to understand
the tame one. As shown by us \cite{AsselmeyerKrol2009}, wild embeddings
are needed to understand exotic smoothness. As explained in \cite{Can:78}
by Cannon, tameness is strongly connected to another topic: decomposition
theory (see the book \cite{Daverman1986}). 

Two embeddings $f,g:N\to M$ are said to be isotopic, if there exists
a homeomorphism $F:M\times[0,1]\to M\times[0,1]$ such that 
\begin{enumerate}
\item $F(y,0)=(y,0)$ for each $y\in M$ (i.e. $F(.,0)=id_{M}$)
\item $F(f(x),1)=g(x)$ for each $x\in N$, and
\item $F(M\times\left\{ t\right\} )=M\times\left\{ t\right\} $ for each
$t\in[0,1]$.
\end{enumerate}
If only the first two conditions can be fulfilled then one call it
concordance. Embeddings are usually classified by isotopy. An important
example is the embedding $S^{1}\to\mathbb{R}^{3}$, known as knot,
where different knots are different isotopy classes.

\subsection{Real cohomology classes and wild embeddings}

Wild embeddings are important to understand usual embeddings. Consider
a closed curve in the plane. By common sense, this curve divides the
plane into an interior and an exterior area. The Jordan curve theorem
agrees with that view completely. But what about one dimension higher,
i.e. consider the embedding $S^{2}\to\mathbb{R}^{3}$? Alexander was
the first who constructed a counterexample, Alexanders horned sphere
\cite{Alex:24}, as wild embedding $I:D^{3}\to\mathbb{R}^{3}$. The
main property of this wild object $D_{W}^{3}=I(D^{3})$ is the non-simple
connected complement $\mathbb{R}^{3}\setminus D_{W}^{3}$. This property
is a crucial point of the following discussion. Given an embedding
$I:D^{3}\to\mathbb{R}^{3}$ which induces a decomposition $\mathbb{R}^{3}=I(D^{3})\cup(\mathbb{R}^{3}\setminus I(D^{3}))$.
In case, the embedding is tame, the image $I(D^{3})$ is given by
a finite complex and every part of the decomposition is contractable,
i.e. especially $\pi_{1}(\mathbb{R}^{3}\setminus I(D^{3}))=0$. For
a wild embedding, $I(D^{3})$ is an infinite complex (but contractable).
The complement $\mathbb{R}^{3}\setminus I(D^{3})$ is given by a sequence
of spaces so that $\mathbb{R}^{3}\setminus I(D^{3})$ is non-simple
connected (otherwise the embedding must be tame) having the homology
of a point (that is true for every embedding). Especially $\pi_{1}(\mathbb{R}^{3}\setminus I(D^{3}))$
is non-trivial whereas its abelization $H_{1}(\mathbb{R}^{3}\setminus I(D^{3}))=0$
vanishes. Therefore $\pi_{1}$ is generated by the commutator subgroup
$[\pi_{1},\pi_{1}]$ with $[a,b]=aba^{-1}b^{-1}$ for two elements
$a,b\in\pi_{1}$, i.e. $\pi_{1}$ is a perfect group.

In the following we will concentrate on wild embeddings of spheres
$S^{n}$ into spheres $S^{m}$ equivalent to embeddings of $\mathbb{R}^{n}$
into $\mathbb{R}^{m}$relative to the infinity $\infty$ point or
to relative embeddings of $D^{n}$ into $D^{m}$ (relative to its
boundary). From the physical point of view, in the case of flat D-branes
when B-field is trivial, branes are seen as topological objects of
a trivial type like $\mathbb{R}^{n},S^{n}$ or $D^{n}$. Lets start
with the case of a finite $k-$dimensional polyhedron $K^{k}$ (i.e.
a piecewise-linear version of a $k-$disk $D^{k}$). Consider the
wild embedding $i:K\to S^{n}$ with $0\leq k\leq n-3$ and $n\geq7$.
Then, as shown in \cite{FerryPedersenVogel1989}, the complement $S^{n}\setminus i(K)$
is non-simple connected with a countable generated (but not finitely
presented) fundamental group $\pi_{1}(S^{n}\setminus i(K))=\pi$.
Furthermore, the group $\pi$ is perfect (i.e. generated by the commutator
subgroup $[\pi,\pi]=\pi$ implying $H_{1}(\pi)=0$) and $H_{2}(\pi)=0$
($\pi$ is called a superperfect group). With other words, $\pi$
is a group where every element $x\in\pi$ can be generated by a commutator
$x=[a,b]=aba^{-1}b^{-1}$ (including the trivial case $x=a,\: b=e$).
By using geometric group theory, we can represent $\pi$ by a grope
(or generalized disk, see Cannon \cite{Can:79}), i.e. a hierarchical
object with the same fundamental group as $\pi$ (see the next subsection).
In \cite{AsselmeyerKrol2009}, the grope was used to construct a non-trivial
involution of the 3-sphere connected with a codimension-1 foliation
of the 3-sphere classified by the real cohomology classes $H^{3}(S^{3},\mathbb{R})$.
By using the suspension \[
\Sigma X=X\times[0,1]/(X\times\left\{ 0\right\} \cup X\times\left\{ 1\right\} \cup\left\{ x_{0}\right\} \times[0,1])\]
of a topological space $(X,x_{0})$ with base point $x_{0}$, we have
an isomorphism of cohomology groups $H^{n}(S^{n})=H^{n+1}(\Sigma S^{n})$.
Thus the class in $H^{3}(S^{3},\mathbb{R})$ induces classes in $H^{n}(S^{n},\mathbb{R})$
for $n>3$ represented by a wild embedding $i:K\to S^{n}$ for some
$k-$dimensional polyhedron. Then small exotic $\mathbb{R}^{4}$ determines
also wild embeddings in higher dimensions, hence higher real cohomology
classes of $n$-spheres:

\begin{theorem}

Let $\mathbb{R}_{H}^{4}$ be some exotic $\mathbb{R}^{4}$ determined
by element in $H^{3}(S^{3},\mathbb{R})$, i.e. by a codimension-1
foliation on the boundary $\partial A$ of the Akbulut cork $A$.
Each wild embedding $i:K^{3}\to S^{p}$ for $p>6$ of a 3-dimensional
polyhedron (as part of $S^{3}$) determines a class in $H^{n}(S^{n},\mathbb{R})$
which represents a wild embedding $i:K^{p}\to S^{n}$ of a $p$ -polyhedron
into $S^{n}$.

\end{theorem}

Now we consider a class of topological quantum D$p$-branes as these
branes which are determined by the wild embeddings $i:K^{p}\to S^{n}$
as above and in the classical and flat limit correspond to tame embeddings.
The quantum character of such D$p$-branes is driven by the presence
of the non-trivial $B$- field which here is encoded in the wild embeddings
$i:K^{3}\to S^{p}$. This last in turn is derived from exotic $\mathbb{R}^{4}$
and is generated by the wild embedding $S^{2}\to S^{3}$\cite{AsselmeyerKrol2009,AsselmeyerKrol2011}.
In the next subsections we will examine the quantum character of wild
embeddings and see how this is related to the class of quantum D-branes
we deal with. Next we will show, directly from the action for such
branes, how the tame embedding emerges in the classical limit.

\subsection{$C^{*}-$algebras associated to wild embeddings}

\begin{figure}
\begin{center}

\includegraphics[height=8cm]{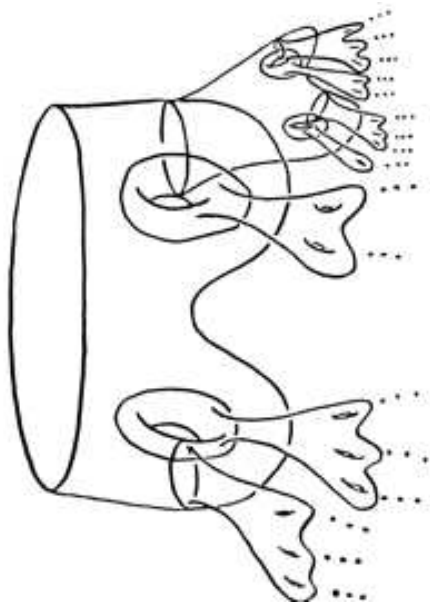}

\end{center}\caption{An example of a grope\label{fig:grope}}

\end{figure}

As described above, a wild embedding $j:K\to S^{n}$ of a polyhedron
$K$ is characterized by its complement $M(K,j)=S^{n}\setminus j(K)$
which is non-simply connected (i.e. the fundamental group $\pi_{1}(M(K,j))$
is non-trivial). The fundamental group $\pi_{1}(M(K,j))=\pi$ of the
complement $M(K,j)$ is a superperfect group, i.e. $\pi$ is identical
to its commutator subgroup $\pi=[\pi,\pi]$ (then $H_{1}(\pi)=0)$
and $H_{2}(\pi)=0$. This group is not finite in case of a wild embedding.
Here we use gropes to represent $\pi$ geometrically. The idea behind
that approach is very simple: the fundamental group of the 2-diemsional
torus $T^{2}$ is the abelian group $\pi_{1}(T^{2})=\left\langle a,b\:|\:[a,b]=aba^{-1}b^{-1}=e\right\rangle =\mathbb{Z}\oplus\mathbb{Z}$
generated by the two standard slopes $a,b$ corresponding to the commuting
generators of $\pi_{1}(T^{2})$. The capped torus $T^{2}\setminus D^{2}$
has an additional element $c$ in the fundamental group generated
by the boundary $\partial(T^{2}\setminus D^{2})=S^{1}$. This element
is represented by the commutator $c=[a,b]$. In case of our superperfect
group, we have the same problem: every element $c$ is generated by
the commuator $[a,b]$ of two other elements $a,b$ which are also
represented by commutators etc. Thus one obtains a hierarchical object,
a generalized 2-disk or a grope (see Fig. \ref{fig:grope}).%
 Now we describe two ways to associate a $C^{*}-$algebra to this
grope. This first approach uses a combination of our previous papers
\cite{AsselmeyerKrol2009,AsselmeyerKrol2010}. Then every grope determines
a codimension-1 foliation of the 3-sphere and vice versa. The leaf-space
of this foliation is a factor $I\! I\! I_{1}$ von Neumann algebra
and we have a $C^{*}-$algebra for the holonomy groupoid. For later
usage, we need a more direct way to construct a $C^{*}-$algebra from
a wild embedding or grope. The main ingredient is the superperfect
group $\pi$, countable generated but not finitely presented group
$\pi$. 

Given a grope $\mathcal{G}$ representing via $\pi_{1}(\mathcal{G})=\pi$
the (superperfect) group $\pi$. Now we define the $C^{*}-$algebra
$C^{*}(\mathcal{G},\pi$) associated to the grope $\mathcal{G}$ with
group $\pi$. The basic elements of this algebra are smooth half-densities
with compact supports on $\mathcal{G}$, $f\in C_{c}^{\infty}(\mathcal{G},\Omega^{1/2})$,
where $\Omega_{\gamma}^{1/2}$ for $\gamma\in\pi$ is the one-dimensional
complex vector space of maps from the exterior power $\Lambda^{2}L$
, of the union of levels $L$ representing $\gamma$ to $\mathbb{C}$
such that \[
\rho(\lambda\nu)=|\lambda|^{1/2}\rho(\nu)\qquad\forall\nu\in\Lambda^{2}L,\lambda\in\mathbb{R}\:.\]
For $f,g\in C_{c}^{\infty}(\mathcal{G},\Omega^{1/2})$, the convolution
product $f*g$ is given by the equality\[
(f*g)(\gamma)=\intop_{[\gamma_{1},\gamma_{2}]=\gamma}f(\gamma_{1})g(\gamma_{2})\]
Then we define via $f^{*}(\gamma)=\overline{f(\gamma^{-1})}$ a $*$operation
making $C_{c}^{\infty}(\mathcal{G},\Omega^{1/2})$ into a $*$algebra.
For each capped torus $T$ in some level of the grope $\mathcal{G}$,
one has a natural representation of $C_{c}^{\infty}(\mathcal{G},\Omega^{1/2})$
on the $L^{2}$ space over $T$. Then one defines the representation\[
(\pi_{x}(f)\xi)(\gamma)=\intop_{[\gamma_{1},\gamma_{2}]=\gamma}f(\gamma_{1})\xi(\gamma_{2})\qquad\forall\xi\in L^{2}(T).\]
The completion of $C_{c}^{\infty}(\mathcal{G},\Omega^{1/2})$ with
respect to the norm \[
||f||=\sup_{x\in M}||\pi_{x}(f)||\]
makes it into a $C^{*}$algebra $C_{c}^{\infty}(\mathcal{G},\pi$).
Finally we are able to define the $C^{*}-$algebra associated to the
wild embedding: \begin{definition} Let $j:K\to S^{n}$ be a wild
embedding with $\pi=\pi_{1}(S^{n}\setminus j(K))$ as fundamental
group of the complement $M(K,j)=S^{n}\setminus j(K)$. The $C^{*}-$algebra
$C_{c}^{\infty}(K,j)$ associated to the wild embedding is defined
to be $C_{c}^{\infty}(K,j)=C_{c}^{\infty}(\mathcal{G},\pi)$ the $C^{*}-$algebra
of the grope $\mathcal{G}$ with group $\pi$. \end{definition}

To get an impression of this superperfect group $\pi$, we consider
a representation $\pi\to G$ in some infinite group. As the obvious
example for $G$ we choose the infinite union $GL(\mathbb{C})=\bigcup_{\infty}GL(n,\mathbb{C})$
of complex, linear groups (induced from the embedding $GL(n,\mathbb{C})\to GL(n+1,\mathbb{C})$
by an inductive limes process). Then we have a homomorphism\[
U:\pi\to GL(\mathbb{C})\]
mapping a commutator $[a,b]\in\pi$ to $U([a,b])\in[GL(\mathbb{C}),GL(\mathbb{C})]$
into the commutator subgroup of $GL(\mathbb{C})$. But every element
in $\pi$ is generated by a commuator, i.e. we have\[
U:\pi\to[GL(\mathbb{C}),GL(\mathbb{C})]\]
and we are faced with the problem to determine this commutator subgroup.
Actually, one has Whitehead's lemma (see \cite{Ros:94}) which determines
this subgroup to be the group of elementary matrices $E(\mathbb{C})$.
One defines the elementary matrix $e_{ij}(a)$ in $E(n,\mathbb{C})$
to be the $(n\times n)$ matrix with 1's on the diagonal,
with the complex number $a\in\mathbb{C}$ in the $(i,j)-$slot, and
0's elsewhere. Analogously, $E(\mathbb{C})$ is the
infinite union $E(\mathbb{C})=\bigcup_{\infty}E(n,\mathbb{C})$. Thus,
every homomorphism descends to a homomorphism\[
U:\pi\to E(\mathbb{C})=[GL(\mathbb{C}),GL(\mathbb{C})]\quad.\]
By using the relation\[
[e_{ij}(a),e_{jk}(b)]=e_{ij}(a)e_{jk}(b)e_{ij}(a)^{-1}e_{jk}(b)^{-1}=e_{ik}(ab)\quad i,j,k\:\mbox{distinct}\]
one can split every element in $E(\mathbb{C})$ into a (group) commutator
of two other elements. By the representation $U:\pi\to E(\mathbb{C})$,
we get a homomorphism of $C_{c}^{\infty}(\mathcal{G},\pi$) into the
usual convolution algebra $C^{*}(E(\mathbb{C}))$ of the group $E(\mathbb{C})$
used later to construct the action of the quantum D-brane.

\subsection{Isotopy classes of wild embeddings and KK theory}

In section \ref{sub:Wild-and-tame-embed} we introduce the notion
of isotopy classes for embeddings. Given two embeddings $f,g:N\to M$
with a special map $F:M\times[0,1]\to M\times[0,1]$ as deformation
of $f$ into $g$, then both embeddings are isotopic to each other.
The definition is independent of the tameness oder wilderness for
the embedding. Now we specialize to our case of wild embeddings $f,g:K\to S^{n}$
with complements $M(K,f)$ and $M(K,g)$. The map $F:S^{n}\times[0,1]\to S^{n}\times[0,1]$
induces a homotopy of the complements $M(K,f)\simeq M(K,g)$ giving
an isomorphism of the fundamental groups $\pi_{1}(M(K,g))=\pi_{1}(M(K,f))$.
Thus, the isotopy class of the wild embedding $f$ is completely determined
by the $M(K,f)$ up to homotopy. Using Connes work on operator algebras
of foliation, our construction of the $C^{*}-$algebra for a wild
embedding is functorial, i.e. an isotopy of the embeddings induces
an isomorphism between the corresponding $C^{*}-$algebras. Given
two non-isotopic, wild embeddings then we have a homomorphism between
the $C^{*}-$algebras only. But every homomorphism (which is not a
isomorphism) between $C^{*}-$algebras $A,B$ gives an element of
$KK(A,B)$ and vice versa. Thus,

\begin{theorem}

Let $j:K\to S^{n}$ be a wild embedding with $\pi=\pi_{1}(S^{n}\setminus j(K))$
as fundamental group of the complement $M(K,j)=S^{n}\setminus j(K)$
and $C^{*}-$algebra $C_{c}^{\infty}(K,j)$. Given another wild embedding
$i$ with $C^{*}-$algebra $C_{c}^{\infty}(K,i)$. The elements of
$KK(C_{c}^{\infty}(K,j),C_{c}^{\infty}(K,i))$ are the isotopy classes
of the wild embedding $j$ relative to $i$. 

\end{theorem}

\section{Wild embeddings as quantum D-branes}

Given a wild embedding $f:K\to S^{n}$ with $C^{*}-$algebra $C^{*}(K,f)$
and group $\pi=\pi_{1}(S^{n}\setminus f(K))$. In this section we
will derive an action for this embedding to derive the D-brane action
in the classical limit. The starting point is our remark above (see
section \ref{sec:Small-exotic-R4-gropes}) that the group $\pi$ can
be geometrically constructed by using a grope $\mathcal{G}$ with
$\pi=\pi_{1}(\mathcal{G})$. This grope was used to construct a codimension-1
foliation on the 3-sphere classified by the Godbillon-Vey invariant.
This class can be seen as element of $H^{3}(BG,\mathbb{R})$ with
the holonomy groupoid $G$ of the foliation. The strong relation between
the grope $\mathcal{G}$ and the foliation gives an isomorphism for
the $C^{*}-$algebra which can be easily verified by using the definitions
of both algebras. As shown by Connes \cite{Connes1984,Connes94},
the Godbillon-Vey class $GV$ can be expressed as cyclic cohomology
class (the so-called flow of weights)\[
GV_{HC}\in HC^{2}(C_{c}^{\infty}(G))\simeq HC^{2}(C_{c}^{\infty}(\mathcal{G},\pi))\]
of the $C^{*}-$algebra for the foliation isomorphic to the $C^{*}-$algebra
for the grope $\mathcal{G}$. Then we define an expression\[
S=Tr_{\omega}\left(GV_{HC}\right)\]
uniquely associated to the wild embedding ($Tr_{\omega}$ is the Dixmier
trace). $S$ is the action of the embedding. Because of the invariance
for the class $GV_{HC}$, the variation of $S$ vanishes if the map
$f$ is a wild embedding. But this expression is not satisfactory
and cannot be used to get the classical limit. For that purpose we
consider the representation of the group $\pi$ into the group $E(\mathbb{C})$
of elementary matrices. As mentioned above, $\pi$ is countable generated
and the generators can be arranged in the embeddings space. Then we
obtain matrix-valued functions $X^{\mu}\in C_{c}^{\infty}(E(\mathbb{C}))$
as the image of the generators of $\pi$ w.r.t. the representation
$\pi\to E(\mathbb{C})$ labelled by the dimension $\mu=1,\ldots,n$
of the embedding space $S^{n}$. Via the representation $\iota:\pi\to E(\mathbb{C})$,
we obtain a cyclic cocycle in $HC^{2}(C_{c}^{\infty}(E(\mathbb{C}))$
generated by a suitable Fredholm operator $F$. Here we use the standard
choice $F=D|D|^{-1}$ with the Dirac operator $D$ acting on the functions
in $C_{c}^{\infty}(E(\mathbb{C}))$. Then the cocycle in $HC^{2}(C_{c}^{\infty}(E(\mathbb{C}))$
can be expressed by\[
\iota_{*}GV_{HC}=\eta_{\mu\nu}[F,X^{\mu}][F,X^{\nu}]\]
using a metric $\eta_{\mu\nu}$in $S^{n}$ via the pull-back using
the representation $\iota:\pi\to E(\mathbb{C})$. Finally we obtain
the action\begin{equation}
S=Tr_{\omega}([F,X^{\mu}][F,X_{\mu}])=Tr_{\omega}([D,X^{\mu}][D,X_{\mu}]|D|^{-2})\label{eq:quantum-D-brane-action}\end{equation}
which can be evaluated by using the heat-kernel of the Dirac operator
$D$.

\subsection{The classical limit}

Similar to the case of the von Neumann algebra of a foliation, the
non-commutativity of the $C^{*}-$algebra $C^{*}(K,f)$ is induced
by the wild embedding $f:K\to S^{n}$. The complexity of the group
$\pi=\pi_{1}(S^{n}\setminus f(K))$ is related to the complexity of
the $C^{*}-$algebra constructed above. Therefore a tame embedding
has a trivial group $\pi$ and we obtain for the $C^{*}-$algebra
$C_{c}^{\infty}(K,f)=\mathbb{C}$, i.e. every operator is a multiplication
operator (multiplication with a complex number). 

From the physical point of view, the non-triviality of the $C^{*}-$algebra
has an interpretation (via the GNS representation) as the observables
algebra of a quantum system. In our case, the non-triviality of the
$C^{*}-$algebra is connected with the wildness of the embedding or
the wild embedding is connected with a quantum system. But then the
classical limit is equivalent to choose a tame embedding $f:K\to S^{n}$
of a $p-$dimensional complex $K$. The Dirac operator $D$ on $K$
acts on usual square-integrable functions, so that $[D,X^{\mu}]=dX^{\mu}$
is finite. The action (\ref{eq:quantum-D-brane-action}) reduces to
\[
S=Tr_{\omega}(\eta_{\mu\nu}(\partial_{k}X^{\mu}\partial^{k}X^{\nu})|D|^{-2})\]
where $\mu,\nu=1,\ldots,n$ is the index for the coordinates on $S^{n}$
and $k=1,\ldots,p$ represents the index of the complex. From the
physical point we expect to obtain an action which describes the embedding
of the brane. For that purpose, we will choose a small fluctuation
$\xi^{k}$ of a fixed embedding given by $X^{\mu}=(x^{k}+\xi^{\mu})\delta_{k}^{\mu}$
with $\partial_{l}x^{k}=\delta_{l}^{k}$. Then we obtain\[
\partial_{k}X^{\mu}\partial^{k}X^{\nu}=\delta_{k}^{\mu}\delta_{k}^{\nu}(1+\partial_{k}\xi^{\mu})(1+\partial_{k}\xi^{\nu})\]
and we use a standard argument to neglect the terms linear in $\partial\xi$:
the fluktuation have no prefered direction and therefore only the
square contributes. Then we have \[
S=Tr_{\omega}(\eta_{\mu\nu}(\delta_{k}^{\mu}\delta_{k}^{\nu}+\partial_{k}\xi^{\mu}\partial^{k}\xi^{\nu})|D|^{-2})\]
for the action. By using a result of \cite{Connes94} one obtains
for the Dixmier trace\[
Tr_{\omega}(|D|^{-2})=2\intop_{K}*(\Phi_{1})\]
with the first coefficient $\Phi_{1}$ of the heat kernel expansion
\cite{bvg:90}\[
\Phi_{1}=\frac{1}{6}R\]
and the action simplifies to\[
S=\intop_{K}\left(\eta_{\mu\nu}(\delta_{k}^{\mu}\delta_{k}^{\nu}+\partial_{k}\xi^{\mu}\partial^{k}\xi^{\nu})\frac{1}{3}R\right)dvol(K)\]
for the main contributions where $R$ is the scalar curvature of $K$
(for $p>2$). Usually we can assume a non-vanishing scalar curvature.
Furthermore we can scale the fluctuation to get the action\[
S=\intop_{K}\left(\eta_{\mu\nu}(\partial_{k}\xi^{\mu}\partial^{k}\xi^{\nu}+\Lambda\eta^{\mu\nu}\right)dvol(K)\]
for some number $\Lambda$ proportional to $R$. It is known that
this action agrees with the usual (Born-Infeld) action\[
S=\intop_{K}\sqrt{\det\left(\eta_{\mu\nu}\partial_{k}\xi^{\mu}\partial^{k}\xi^{\nu}\right)}dvol(K)\]
of flat $p-$branes ($p>2)$ for $\Lambda>0$ (i.e.$R>0$) with vanishing
$B-$field Thus we obtain a (quantum) D-brane action by using wild
embeddings for the description of a quantum D-brane and the flux $H$
represented by the wild embedding $S^{2}\to S^{3}$. These data, in
the classical limit, reduce to the BI action for flat D$p$-brane.
We will further investigate this point in a separate paper.

\section{The 4-dimensional origin of quantum D-branes}

The argumentation above can be simply resummed by the following arguments:
\begin{enumerate}
\item Given an embedding $f:K^{p}\to S^{n}$ of a $p-$complex $K$ into
a $n-$sphere. 
\item This embedding is wild, if the complement $S^{n}\setminus f(K)$ is
non-simple connected, i.e. the fundamental group $\pi_{1}(S^{n}\setminus f(K))\not=0$
does not vanish.
\item We define a topological quantum $p-$brane as the wild embedding $f$.
\item The fundamental group $\pi=\pi_{1}(S^{n}\setminus f(K))$ is a perfect
group, i.e. purely generated by the commutators $\pi=[\pi,\pi]$.
\item This group can be geometrically represented by a 2-complex, called
a generalized disk or grope.
\item From this grope $\mathcal{G}$ we constructed a non-trivial $C^{*}-$algebra
$C_{C}^{\infty}(\mathcal{G},\pi)$. 
\item Non-trivial $B$-field $H\in H^{3}(S^{3},\mathbb{Z})$ is represented
by the wild embedding $S^{2}\to S^{3}$.
\end{enumerate}
The grope is a 2-complex sometimes equipped with an embedding into
the Euclidean space $\mathbb{E}^{3}$. As shown in \cite{AsselmeyerKrol2009},
one can also use it to describe small exotic $\mathbb{R}^{4}$'s (see
also some details in section \ref{sec:Small-exotic-R4-gropes}). At
the first view we have two possible interpretations, the 2-dimensional
grope and the 4-dimensional exotic $\mathbb{R}^{4}$, which are rather
independent of each other. But in the derivation of the action above,
we used implicitly the result that an exotic $\mathbb{R}^{4}$ (and
the grope constructed from it) is (partly) classified by the Godbillon-Vey
invariant. Therefore our topological quantum D-brane is generated
by a small exotic $\mathbb{R}^{4}$ too.

\section{Conclusions}

Every small exotic $\mathbb{R}^{4}$ is a very rich many-facets hybrid
object which links, among others, $C^{\star}$ convolution algebras,
K-theory, foliations and topology in particular. It can also be represented
by a wild embedding $S^{2}\hookrightarrow S^{3}$ \cite{AsselmeyerKrol2009}.
When $\mathbb{R}^{4}$ is taken with its standard smooth structure,
hence smoothness agrees with product topology, then all complexities
of the structures disappear. In this paper we argue that exotic $\mathbb{R}^{4}$'s
are involved in the formalism of string theory also at the non-perturbative
domains where branes are considered as quantum objects. Especially,
exotic $\mathbb{R}^{4}$'s determine a class of topological quantum
D$p$-branes. On the other hand the persented results support our
conjecture from \cite{AsselmeyerKrol2010}, stating that:\\
\emph{The exotic small $\mathbb{R}^{4}$ lies at the heart of
quantum gravity in dimension 4. Especially it is a quantized object.}\\
The connections between 4-exotics and NS and D-branes in various
string backgrounds were given in \cite{AsselmeyerKrol2011} and then
extended formally to the quantum regime of D-branes \cite{AsselmeyerKrol2011b}.
Here we further extend this relation and propose a topological mechanism
generating classes of branes and charges in some backgrounds. We study
the case of quantum D-branes using $C^{*}-$algebras. The topological
mechanism behind quantum branes is the wild embedding of 2-spheres
into $S^{3}$ as well $S^{3}$ into higher dimensional spheres. These
last embeddings generate D-branes which are considered as \emph{topological
quantum D-branes} whereas the non-trivial class, $H$, or $B$-field,
is derived from the first wild embedding, i.e. $S^{2}\to S^{3}$.
The presented mechanism generates quantum topological D$p$-branes
when the non-trivial $B$-field on $S^{3}$ is given as a (quantum)
wild embedding. On the other hand classical branes are considered
as submanifolds or K-homology cycles. In case of the quantum regime
they are usually described as K-theory classes on separable $C^{\star}$-algebras
\cite{Szabo2008a}. It appears that many kinds of this $C^{\star}$-algebraic
presentations have, in turn, topological origins and are again derived
from the wild embeddings. 

Taking the classical limit of such quantum D$p$-branes, where $B$-field
is confined on $S^{3}\subset{\rm WV}({\rm D}p)$ corresponding to
wild embeddings, one gets tame and flat embeddings of $p$-complexes.
This follows in particular from the reduction of the quantum action
to BI action. The results can be roughly summarized by:\\
\emph{The exotic small $\mathbb{R}^{4}$ as described by codimension-1
foliations on the 3-sphere is the germ of wide range of effects on
D-branes. A }topological\emph{ quantum Dp-brane is related to a wild
embedding of a $p-$dimensional complex into a $n-$dimensional space
described by a two-dimensional complex, a grope.} \emph{The grope
is the main structure to get the relation between the exotic small
$\mathbb{R}^{4}$ and the codimension-1 foliation on the 3-sphere
\cite{AsselmeyerKrol2009,AsselmeyerKrol2011,AsselmeyerKrol2011b,AsselmeyerKrol2011a}.}
\\
 The description of the wild embedding is rather independent of
the dimension ($n>6$, $p>2$) which is the reason why small exotic
$\mathbb{R}^{4}$'s appear in different dimensions as germs of higher
dimensional topological quantum branes.

\section*{Acknowledgment}

T.A. wants to thank C.H. Brans and H. Ros\'e for numerous discussions
over the years about the relation of exotic smoothness to physics.
J.K. benefited much from the explanations given to him by Robert Gompf
regarding 4-smoothness several years ago, and discussions with Jan
S{\l}adkowski.


\end{document}